%Paper: hep-ph/9311379
%From: BILL KILGORE <KILGORE@theor3.lbl.gov>
%Date: Tue, 30 Nov 1993 14:49:48 -0800 (PST)

%%%%%%%%%%%%%%%%%%%%%%%%%%%%%%%%%%%%%%%%%%%%%%%%%%%%%%%%%%%%%%%%%%%
%                       INSTRUCTIONS
%
% This paper has 2 postscript figures which have been included as
% a uuencoded compressed tar file with instructions for unpacking,
% using the shell script uufiles (which is available from the
% bulletin board).  The figures are not imbedded in TeX and must be
% printed separately.
%
%%%%%%%%%%%%%%%%%%%%%%%%%%%%%%%%%%%%%%%%%%%%%%%%%%%%%%%%%%%%%%%%%%%
\input harvmac
%%%%%%%%%%%%%%%%%%%%%%%%%%%%%%%%%%%%%%%%%%%%%%%%%%%%%%%%%%%%%%%%%%%
%  Customizations of Harvmac
%%%%%%%%%%%%%%%%%%%%%%%%%%%%%%%%%%%%%%%%%%%%%%%%%%%%%%%%%%%%%%%%%%%
\catcode`\@=11 % This allows us to modify PLAIN macros.
\ifx\answ\bigans \else
\output={\ifnum\pageno<1 %
  \shipout\vbox{\speclscape{\hsize\fullhsize\makeheadline}
    \hbox to \fullhsize{\hfill\pagebody\hfill}
    \speclscape{\hsize\fullhsize\makefootline}}
\advancepageno
  \else
  \almostshipout{\leftline{\vbox{\pagebody\makefootline}}}\advancepageno 
  \fi}
\def\almostshipout#1{\if L\l@r \count1=1 \message{[\the\count0.\the\count1]}
      \global\setbox\leftpage=#1 \global\let\l@r=R
 \else \count1=2
  \shipout\vbox{\speclscape{\hsize\fullhsize\makeheadline}
      \hbox to\fullhsize{\box\leftpage\hfil#1}}  \global\let\l@r=L\fi}
\fi
\catcode`\@=12 % @ is no longer a letter
\def\Title#1#2{\abstractfont\hsize=\hstitle{#1}%
\vskip .5in\centerline{\titlefont #2}\abstractfont\vskip .5in\pageno=0
\footline={\hss}}
\def\Date#1{\vfill\leftline{#1}\tenpoint\supereject\global\hsize=\hsbody%
\pageno=1\footline={\hss\tenrm\folio\hss}}% 	restores pagenumbers
%  Footnotes
\def\footend{\def\foot{\global\advance\ftno by1\chardef\wfile=\ftfile
$^{\the\ftno}$\ifnum\ftno=1\immediate\openout\ftfile=\jobname.foot\fi%
\immediate\write\ftfile{\noexpand\smallskip%
\noexpand\item{f\the\ftno:\ }\pctsign}\findarg}%
\def\footatend{\vfill\eject\immediate\closeout\ftfile{\parindent=20pt
\centerline{\bf Footnotes}\nobreak\bigskip\input \jobname.foot }}}
%  References
\def\nref#1{\xdef#1{[\the\refno]}\writedef{#1\leftbracket#1}%
\ifnum\refno=1\immediate\openout\rfile=\jobname.refs\fi
\global\advance\refno by1\chardef\wfile=\rfile\immediate
\write\rfile{\noexpand\item{#1\ }\reflabeL{#1\hskip.31in}\pctsign}\findarg}
\def\listrefs{\footatend\vfill\supereject\immediate\closeout\rfile\writestoppt
\baselineskip=14pt\centerline{{\bf References}}\bigskip{\frenchspacing%
\parindent=20pt\escapechar=` \input \jobname.refs\vfill\eject}\nonfrenchspacing}
\def\startrefs#1{\immediate\openout\rfile=\jobname.refs\refno=#1}
%  Figures

%
\def\fig{\the\figno\nfig}
\def\nfig#1{\xdef#1{\the\figno}%
\writedef{#1\leftbracket \noexpand~\the\figno}%
\ifnum\figno=1\immediate\openout\ffile=\jobname.figs\fi\chardef\wfile=\ffile%
\immediate\write\ffile{\noexpand\medskip\noexpand\item{Figure\ \the\figno. }
\reflabeL{#1\hskip.55in}\pctsign}\global\advance\figno by1\findarg}
\def\listfigs{\vfill\eject\immediate\closeout\ffile{\parindent40pt
\baselineskip14pt\centerline{{\bf Figure Captions}}\nobreak\medskip
\escapechar=` \input \jobname.figs\vfill\eject}}
\def\xfig{\expandafter\xf@g}\def\xf@g fig.\penalty\@M\ {}
\def\figs#1{figs.~\f@gs #1{\hbox{}}}
\def\f@gs#1{\edef\next{#1}\ifx\next\em@rk\def\next{}\else
\ifx\next#1\xfig #1\else#1\fi\let\next=\f@gs\fi\next}
%  Labels
\def\writedefs{\immediate\openout\lfile=\jobname.lbls \def\writedef##1{%
\immediate\write\lfile{\string\def\string##1\rightbracket}}}
%  Table of Contents
\def\writetoc{\immediate\openout\tfile=\jobname.tocs 
   \def\writetoca##1{{\edef\next{\write\tfile{\noindent ##1 
   \string\leaderfill {\noexpand\number\pageno} \par}}\next}}}
%       and this lists table of contents on second pass
\def\listtoc{\centerline{\bf Contents}\nobreak\medskip{\baselineskip=12pt
 \parskip=0pt\catcode`\@=11 \input \jobname.tocs \catcode`\@=12 \bigbreak\bigskip}}
%  Title Fonts
\edef\tfontsize{\ifx\answ\bigans scaled\magstep1\else scaled\magstep1\fi}
\font\titlerm=cmbx10 \tfontsize \font\titlerms=cmbx7 \tfontsize
\font\titlermss=cmbx5 \tfontsize \font\titlei=cmmib10 \tfontsize
\font\titleis=cmmi7 \tfontsize \font\titleiss=cmmi5 \tfontsize
\font\titlesy=cmbsy10 \tfontsize \font\titlesys=cmsy7 \tfontsize
\font\titlesyss=cmsy5 \tfontsize \font\titleit=cmbxti10 \tfontsize
\skewchar\titlei='177 \skewchar\titleis='177 \skewchar\titleiss='177
\skewchar\titlesy='60 \skewchar\titlesys='60 \skewchar\titlesyss='60
\def\titlefont{\def\rm{\fam0\titlerm}% switch to title font
\textfont0=\titlerm \scriptfont0=\titlerms \scriptscriptfont0=\titlermss
\textfont1=\titlei \scriptfont1=\titleis \scriptscriptfont1=\titleiss
\textfont2=\titlesy \scriptfont2=\titlesys \scriptscriptfont2=\titlesyss
\textfont\itfam=\titleit \def\it{\fam\itfam\titleit}\rm}
%
%%%%%%%%%%%%%%%%%%%%%%%%%%%%%%%%%%%%%%%%%%%%%%%%%%%%%%%%%%%%%%%%%%%
%  Personal Macros
%%%%%%%%%%%%%%%%%%%%%%%%%%%%%%%%%%%%%%%%%%%%%%%%%%%%%%%%%%%%%%%%%%%
%
%%% Boldgreek.tex
% Define a new family: see The TeXbook, Exercise 17.20 and its answer.
\font\xiibifull=cmmib10 % bold math italic
\font\xiibimed=cmmib10 scaled 800
\font\xiibismall=cmmib10 scaled 640
\newfam\boldgreekfam  %
\textfont\boldgreekfam=\xiibifull \scriptfont\boldgreekfam=\xiibimed
\scriptscriptfont\boldgreekfam=\xiibismall
% The bold versions of the lower-case Greek letters.
\def\boldxi{{\fam\boldgreekfam\mathchar"7118 } }
\def\boldpi{{\fam\boldgreekfam\mathchar"7119 } }
\def\boldsigma{{\fam\boldgreekfam\mathchar"711B } }
\def\boldSigma{{\fam=6{\mathchar"7006 } }}

\def\ie{\hbox{\it i.e.}}

\def\dm{\partial^{\ }_{\mu}}  \def\dn{\partial^{\ }_{\nu}}
\def\calDsl{\,\,\raise.15ex\hbox{/}\mkern-13.5mu {\cal D}}
\def\calAsl{\,\,\raise.15ex\hbox{/}\mkern-14.5mu {\cal A}}
\def\epsl{\lower.1ex\hbox{/}\mkern-8.5mu \epsilon}
\def\psl{\lower.3ex\hbox{/}\mkern-9mu p}

\def\Lm{{\bf L}_{\mu}}            \def\Ln{{\bf L}_{\nu}}
\def\Lr{{\bf L}_{\rho}}           \def\Ls{{\bf L}_{\sigma}}
\def\gmma{\raise .25ex \hbox{$\scriptstyle\gamma$}}

\def\title{Anomalous Condensates and the Equivalence Theorem}
\def\lblnum{LBL-34900}
\setbox254=\hbox{hep-ph/9311379}
\dimen254=\hstitle \advance\dimen254 by -\wd254
\def\topline{\halign{\hbox to \dimen254{##\hfil}&##\hfil \cr
    30 November 1993 & \lblnum\cr &\box254\cr}}

\font\smallit=cmti10 scaled 800

\def\DOE{This work was supported by the Director, Office of Energy 
research, Office of High Energy and Nuclear Physics, Division of High 
Energy Physics of the U.S. Department of Energy under Contract 
DE-AC03-76SF00098.}

\def\abstract{A recently published report has called into question the
validity of the equivalence theorem in dynamically broken gauge theories
in which the fermions making up the symmetry breaking condensate lie in an
anomalous representation of the broken gauge group.  Such a situation can
occur if the gauge anomaly is cancelled by another sector of the theory.
Using the example of the one family Standard Model without scalar Higgs
structure, we analyze a low energy effective theory which preserves the
symmetries of the fundamental theory and demonstrate the validity of the
equivalence theorem in this class of models.}

\Title{\topline}{\title\footnote{${}^{\star}$}\DOE}

\centerline{William B. Kilgore}

\vskip .5in

{\it
\centerline{     Theoretical Physics Group}
\centerline{     Lawrence Berkeley Laboratory}
\centerline{     and}
\centerline{     Department of Physics}
\centerline{     University of California}
\centerline{     Berkeley, California 94720}
}

\vskip .5in

\abstract

\vfil\eject
\pageno=-2
\hbox to \hstitle{\hfil}
\vskip 1in

\centerline{\bf Disclaimer}

\vskip .2in

{\parindent=.5cm\narrower\smallskip\sevenrm
This document was prepared as an account of work sponsored by the United
States Government.  Neither the United States Government nor any agency
thereof, nor The Regents of the University of California, nor any of their
employees, makes any warranty, express or implied, or assumes any legal
liability or responsibility for the accuracy, completeness, or usefulness
of any information, apparatus, product, or process disclosed, or represents
that its use would not infringe privately owned rights.  Reference herein
to any specific commercial products process, or service by its trade name,
trademark, manufacturer, or otherwise, does not necessarily constitute or
imply its endorsement, recommendation, or favoring by the United States
Government or any agency thereof, or The Regents of the University of
California.  The views and opinions of authors expressed herein do not
necessarily state or reflect those of the United States Government or any
agency thereof of The Regents of the University of California and shall
not be used for advertising or product endorsement purposes.
\smallskip}

\vskip 2in

\centerline{\smallit Lawrence Berkeley Laboratory is an 
  equal opportunity employer.}
\footline={\hfill\tenrm\folio\hfill}
\Date{}

\newsec{Introduction}
Recently, Donoghue and Tandean\ref\DT{
  J.F.~Donoghue and J.~Tandean, Phys. Lett. B 301 (1993) 372.}\ 
have discussed models of
dynamical gauge symmetry breaking in which the fermions that
participate in the symmetry breaking condensate lie in an anomalous
representation of the broken symmetry.  To preserve the gauge symmetry
in the quantized theory there must be no net gauge anomaly.\ref\GJ{
  D.J.~Gross and R.~Jackiw, Phys. Rev. D 6 (1972) 477.
\hfil\break
  \ C.~Bouchiat, J.~Illiopoulos and Ph.~Meyer, Phys. Lett. B 38 (1972)
      519.}\ 
Therefore there must also be additional fermions which cancel the
gauge anomaly but do not take part in the condensate.  They conclude
that the equivalence theorem,\ref\CG{
  M.S.~Chanowitz and M.K.~Gaillard, Nucl. Phys. B 261 (1985) 379.
\hfil\break
  \ G.J.~Gounaris, R.~K\"ogerler and H.~Neufeld, Phys. Rev. D 34 (1986)
      3257.
\hfil\break
  \ J.~Bagger and C.~Schmidt, Phys. Rev. D 41 (1990) 264.
\hfil\break
  \ H.~Veltman, Phys. Rev. D 41 (1990) 2294.
\hfil\break
  \ W.~Kilgore, Phys. Lett. B 294 (1992) 257 ,1992.
\hfil\break
  \ H.J.~He, Y.P.~Kuang and X.~Yi, Phys. Rev. Lett. 69 (1992) 2619.
\hfil\break
  \ H.J.~He, Y.P.~Kuang and X.~Yi, TUIMP preprint TUIMP-TH-92/51}
\ref\CLT{\ifx\answ\bigans J.M.~Cornwall, D.N.~Levin, and G.~Tiktopoulos, Phys. Rev. D 10 (1974)
 1145.
 \else J.M.~Cornwall, D.N.~Levin, and G.~Tiktopoulos,
    \hfil\break\indent Phys. Rev. D 10 (1974) 1145. \fi
\hfil\break
  \ C.E.~Vayonakis, Lett. Nuovo Cim. 17 (1976) 383.
\hfil\break
  \ B.W.~Lee, C.~Quigg and H.~Thacker, Phys. Rev. D 16 (1977) 1519.
\hfil\break
  \ D.~Soper and Z.~Kunst, Nucl. Phys. B 296 (1988) 253.}\ 
 which relates the
scattering amplitudes of high-energy ($E \gg M$) longitudinal gauge
bosons to those of the unphysical would-be Goldstone bosons, is
invalid in such a theory.  Specifically, they find that there is a
non-zero scattering amplitude for the production of would-be Goldstone
bosons via the Wess-Zumino anomaly interaction, while the production
of gauge bosons, longitudinal or otherwise, through fermion triangle
diagrams is forbidden by the anomaly cancellation condition.

This result would be quite disturbing, since the equivalence theorem is
proven\CG\ to follow from the BRS\ref\BRS{
  C.~Becchi, A.~Rouet and R.~Stora, Comm. Math. Phys. B 261 (1975)
      379.}\ 
identities of the theory.\footnote{${}^\star$}{The existing proofs of the
equivalence theorem are all in the context of a Higgs theory, but the
proof can be carried over to a low energy effective field 
theory.\ref\Wka{W.~Kilgore, in preparation.}}  In this paper, we
resolve the apparent paradox by showing that the
anomaly cancellation condition does not imply that the production of
gauge bosons via fermion triangle diagrams must vanish.  Anomaly
cancellation guarantees gauge invariance and causes the
mass-independent pieces of the triangle diagrams to cancel one
another.  The mass dependent pieces are not constrained to cancel and
in general do not cancel except when the fermion masses are
equal.  When the interactions of the would-be Goldstone bosons are
properly  formulated (\ie\ such that the symmetries of the full
theory, such as gauge invariance, are conserved), would-be Goldstone
boson production is consistent with gauge boson production so that the
equivalence theorem is indeed valid in anomalous condensate models.

\newsec{A Toy Model}

Let us carefully construct a simple theory that accomplishes
spontaneous gauge symmetry breaking via an anomalous condensate.  We
consider a theory with the gauge structure of the Standard Model, a
single family of fermions, and nothing else.  In particular, there is
no scalar Higgs sector to break the electroweak symmetry and give
masses to the gauge bosons or the fermions. Thus, the electroweak
symmetry is broken dynamically, as in technicolor models, by the
spontaneous chiral symmetry breakdown of QCD.\ref\WWS{
  M.~Weinstein, Phys. Rev. D 7 (1973) 1854; Phys. Rev. D 8 (1973) 2511.
\hfil\break
  \ S.~Weinberg, Phys. Rev. D 19 (1979) 1277.
\hfil\break
  \ L.~Susskind, Phys. Rev. D 20 (1979) 2619.}\ 

The Lagrangian for our theory is:
\eqn\EQi{\vcenter{\openup2\jot\halign{$\displaystyle\hfil#$
                                &$\displaystyle#\hfil$\cr
 {\cal L} = &-{1\over4}G^{a\mu\nu}G^{a}_{\mu\nu}
    - {1\over4}W^{i\mu\nu}W^{i}_{\mu\nu}
    - {1\over4}B^{\mu\nu}B_{\mu\nu}\cr
   & + \overline q_{L} i\Dsl q_{L} + \overline q_{R} i\Dsl q_{R}
     + \overline l_{L} i\Dsl l_{L} + \overline e_{R} i\Dsl e_{R}
  ,\cr}}}
where $G^{a}_{\mu\nu}$, $W^{i}_{\mu\nu}$ and $B_{\mu\nu}$ are the QCD,
$SU(2)_{L}$, and hypercharge field strength tensors, respectively and
\eqn\EQii{\vcenter{\openup2\jot\halign{$\displaystyle#$\cr
   q=\left(u\atop d\right),\hfil l=\left(\nu\atop e\right),\cr
   D_{\mu} = \dm - ig_{3}G^{a}_{\mu}{\lambda^{a}\over2}
      - ig_{2}W^{i}_{\mu}{\sigma^{i}L\over2}
      - ig_{1}B_{\mu}Y \cr}}}
where $\lambda^{a}$ acts only on the color triplet quarks, $L$ is the
left chirality projection operator, $L = {1-\gamma_{5}\over2}$, and
$Y$ represents the usual Standard Model hypercharge quantum number for
each fermion species.

Being massless, the quarks exhibit an exact $SU(2)_L \otimes SU(2)_R$
chiral symmetry.  At scale $\Lambda_{\chi}$, a fermion condensate
forms with a non-zero vacuum expectation value,
\eqn\EQiii{\langle \overline{u}u + \overline{d}d \rangle \ne 0,}
which breaks the $SU(2)_L \otimes SU(2)_R$ symmetry down to $SU(2)_V$.
Associated with this symmetry breakdown are three Goldstone bosons,
the pions, one for each broken symmetry generator.

The pions transform under a nonlinear representation of $SU(2)_L
\otimes SU(2)_R$, forming a (linear) triplet representation of the
unbroken $SU(2)_V$ symmetry, but transforming nonlinearly under the
broken axial $SU(2)_A.$  All such nonlinear realizations are
physically equivalent\ref\CCWZ{\ifx\answ\bigans
  S.~Coleman, J.~Wess and B.~Zumino, Phys. Rev. 177 (1969) 2239.
\hfil\break
  \ C.~Callan, S.~Coleman, J.~Wess and B.~Zumino, Phys. Rev. 177 (1969)
      2247.\ 
\else
  S.~Coleman, J.~Wess and B.~Zumino, Phys. Rev. 177 (1969) 2239.
\hfil\break
  \ C.~Callan, S.~Coleman, J.~Wess and B.~Zumino, 
\hfil\break\indent
  Phys. Rev. 177 (1969) 2247.\fi}\ 
and their interactions are described
by the so-called chiral Lagrangian, which may be written as
\eqn\EQiv{{\cal L}_{GB} = {F_{\pi}^2\over4} \Tr \left\lgroup
   D^{\mu}\boldSigma^{\dagger} D_{\mu}\boldSigma \right\rgroup + \dots
}\ where
\eqn\EQv{\vcenter{\openup2\jot\halign{$\displaystyle\hfil#\hfil$\cr
    \boldSigma = \exp\left(i{{{\boldsigma}\cdot\boldpi}
          \over F_{\pi}}\right),\cr
    D_{\mu}\boldSigma = \dm\boldSigma
      - ig_{2}W^{i}_{\mu}{\sigma^{i}\over2}\boldSigma
      + ig_{1}B_{\mu}\boldSigma{\sigma^{3}\over2},\cr
}}}
the $\pi^{i}$'s are the pion fields, the $\sigma^{i}$'s are the Pauli
matrices, and $F_{\pi}$, called the pion decay constant, is the
strength of the coupling of the pions to the axial $SU(2)_{A}$
currents,
\eqn\EQvi{\left\langle0\left|J^{i}_{\mu}\right|\pi^{j}(q)\right\rangle =
  iF_{\pi}q_{\mu}\delta^{ij}.}
Under $SU(2)_L\otimes SU(2)_R$ transformations, $\boldSigma$
transforms as
\eqn\EQvii{\boldSigma \longmapsto L\boldSigma R^{\dagger}.}
The ellipses in equation~\EQiv\ indicate terms that involve higher
covariant derivatives and are assumed to be small at low energy.

In addition to pion kinetic terms and multi-pion interaction terms, we
find that ${\cal L}_{GB}$ includes vector boson mass terms and gauge
boson -- pion mixing terms which are exactly those found in the
Standard Model with $F_{\pi}$ taking the place of the Higgs vacuum
expectation value $v$ and the pions serving as the would-be Goldstone
bosons.  Thus the pions are not physical degrees of freedom, but are
absorbed into the gauge bosons via the Higgs mechanism.  Since the
vector  mass terms have changed in scale only ($v \rightarrow
F_{\pi}$), the Weinberg angle, parameterizing the mixing of $W^{3}$ and
$B$ into $\gamma$ and $Z$ as well as the $W$ to $Z$ mass ratio, is the
same as in the Standard Model.

\newsec{Chiral Quarks}

The quarks participate in the condensate and acquire mass through
their coupling to it.  Below the chiral symmetry breaking scale,
$\Lambda_{\chi}$, the quark Lagrangian may be written\ref\MG{
  A.~Manohar and H.~Georgi, Nucl. Phys. B 234 (1984) 189.}
\ref\MM{A.~Manohar and G.~Moore, Nucl. Phys. B 243 (1984) 55.}\ 
as
\eqn\EQviii{{\cal L}_{q} = \overline q_{L} i\Dsl q_{L} + \overline q_{R}
     i\Dsl q_{R} - m_{Q}\left(\overline q_{L}\boldSigma q_{R}
  + \overline q_{R}\boldSigma^{\dagger} q_{L}\right) + \dots
}
Upon expanding $\boldSigma$ in powers of $\pi$, one sees that the
zeroth order term is the mass term for the quarks.  This mass,
$m_{Q}$, is called the constituent mass of the quarks and is expected
to be of the same order of magnitude as the chiral symmetry breaking
scale $\Lambda_{\chi}$.

The Lagrangian in equation~\EQviii\ gives us a perfectly good description
of the low energy quark interactions, but does not explicitly exhibit
decoupling of the massive quarks from very low energy ($s \ll
m_{Q}^{2}$) processes.  For instance, the coupling of two vector
currents to the pseudoscalar pions via quark triangle diagrams (VVP
triangles),  does not vanish for $s \ll m_{Q}^{2}$, even in the limit
that $m_{Q}$ approaches infinity.\ref\JS{
  J.~Steinberger, Phys. Rev. 76 (1949) 1180.}\ 
Gauge invariance forbids quark decoupling; the leptons carry an anomalous
gauge dependence  
which is exactly cancelled by the anomalous gauge dependence of the
quarks.  If the quarks decoupled, there would be nothing to cancel the
gauge dependence of the leptons.  It is possible to write the
Lagrangian in a form such that the quarks can be
decoupled,\ref\DF{
  E.~D'Hoker and E.~Farhi, Nucl. Phys. B 248 (1984) 59,77.}\ 
but there are subtleties involved and we will
examine the procedure carefully.  (We follow here the work of Manohar
and Moore.\MM)

The anomalous gauge dependence of the quark Lagrangian in
equation~\EQviii\ is tied to the fact that the left{-} and right-handed
quarks transform independently under $SU(2)_{L} \otimes SU(2)_{R}$
rotations:
\eqn\EQix{q_{L} \longmapsto L q_{L}, \hskip 2in q_{R} \longmapsto R q_{R}.}
However, it is possible to define a new basis for the quarks so that
both left{-} and right-handed fields transform in the same way.  To
that end, we define the matrix $\boldxi$ as
\eqn\EQx{\boldxi^{2} \equiv \boldSigma.}
Under chiral rotations, $\boldxi$ transforms as
\eqn\EQxi{\boldxi \longmapsto L\boldxi U^{\dagger}(x)
  = U(x)\boldxi R^{\dagger},}
where $U(x)$ is a unitary matrix implicitly defined by the above
equations.  Using $\boldxi$ one may rotate the quarks to a new basis
$Q$, defined by
\eqn\EQxii{Q_{L} = \boldxi^{\dagger}q_{L},
  \qquad\qquad Q_{R} = \boldxi q_{R}}
which transform as
\eqn\EQxiii{Q_{L,R} \longmapsto U(x)Q_{L,R}.}

We define vector and axial vector fields as
\eqn\EQxiv{\ifx\answ\bigans
  \vcenter{\openup2\jot\halign{$\displaystyle\hfil#\hfil$\cr
  {\cal V}_{\mu} = {1\over2}\left(\boldxi(D_{\mu}\boldxi^{\dagger})
   + \boldxi^{\dagger}(D_{\mu}\boldxi) \right),
  \hskip 4em
   {\cal A}_{\mu} = {i\over2}\left(\boldxi(D_{\mu}\boldxi^{\dagger})
    - \boldxi^{\dagger}(D_{\mu}\boldxi) \right),\cr}}
\else
  \vcenter{\openup2\jot\halign{$\displaystyle\hfil#\hfil$\cr
   {\cal V}_{\mu} = {1\over2}\left(\boldxi(D_{\mu}\boldxi^{\dagger})
    + \boldxi^{\dagger}(D_{\mu}\boldxi) \right),\cr 
   {\cal A}_{\mu} = {i\over2}\left(\boldxi(D_{\mu}\boldxi^{\dagger})
    - \boldxi^{\dagger}(D_{\mu}\boldxi) \right),\cr}}\fi
}
where
\eqn\EQxv{\ifx\answ\bigans
  \vcenter{\openup2\jot\halign{$\displaystyle\hfil#\hfil$\cr
    D_{\mu}\boldxi \equiv \dm\boldxi - ig_{2}W^{i}_{\mu}
     {\sigma^{i}\over2}\boldxi -ig_{1}B_{\mu}{1\over6}\boldxi,
   \hskip 4em
    D_{\mu}\boldxi^{\dagger} \equiv \dm\boldxi^{\dagger} - ig_{1}B_{\mu}
	  \left({1\over6}+{\sigma^{3}\over2}\right)\boldxi^{\dagger}.\cr}}
\else
  \vcenter{\openup2\jot\halign{$\displaystyle\hfil#\hfil$\cr
    D_{\mu}\boldxi \equiv \dm\boldxi - ig_{2}W^{i}_{\mu}
     {\sigma^{i}\over2}\boldxi -ig_{1}B_{\mu}{1\over6}\boldxi,\cr
    D_{\mu}\boldxi^{\dagger} \equiv \dm\boldxi^{\dagger} - ig_{1}B_{\mu}
	  \left({1\over6}+{\sigma^{3}\over2}\right)\boldxi^{\dagger}.\cr}}\fi
}
which transform as
\eqn\EQxvi{\vcenter{\openup2\jot\halign{$\displaystyle\hfil#\hfil$\cr
  {\cal V}_{\mu} \longmapsto U{\cal V}_{\mu}U^{\dagger} + U\dm
  U^{\dagger}, \hskip 2cm
  {\cal A}_{\mu} \longmapsto U{\cal A}_{\mu}U^{\dagger}.\cr
}}}
In this notation, the Goldstone boson Lagrangian is written as
\eqn\EQxvii{{\cal L}_{GB} = {F_{\pi}^2\over4} \Tr\left\lgroup {\cal A}^{\mu}
    {\cal A}_{\mu}\right\rgroup + \dots}
and the quark Lagrangian as
$${\cal L}_{Q} = \overline Q\left(i\calDsl^{Q} - m_{Q}\right)Q
   + \overline Q\calAsl\gamma_{5}Q + \dots,$$
where
\eqn\EQxviii{{\cal D}^{Q}_{\mu}Q = \left(\dm - ig_{3}G^{a}_{\mu}{\lambda^{a}\over2}
   + {\cal V}_{\mu}\right)Q.}

The Lagrangian in equation~\EQviii\ and the above expression for ${\cal
L}_{Q}$ are not equivalent.  Since $Q_{L}$ and $Q_{R}$ transform in
the same way, ${\cal L}_{Q}$ is not anomalous under local chiral
$SU(2)_{L} \otimes SU(2)_{R}$ rotations and since electroweak gauge
rotations are simply a subset of these local chiral rotations, ${\cal
L}_{Q}$ has no anomalous gauge dependence.  Somehow, in changing quark
bases, we seem to have lost the anomalous gauge dependence that we
need so that the complete theory is gauge invariant.  This apparent
paradox arises because the change of variables from $q$ to $Q$ changes
the measure of the path integral, requiring a non-trivial Jacobian
factor.  This Jacobian factor carries the anomalous gauge dependence.

A change in the measure of the path integral is difficult to
incorporate in a perturbative expansion.  So, just as we move the
gauge fixing condition and the Fadeev-Popov determinant out of the
measure of the path integral and into the effective Lagrangian, here
too we add a new term to the Lagrangian which exactly compensates for
the change in measure induced by the quark rotation.  This
compensating term is the Wess-Zumino anomaly term,\ref\WZ{
  J.~Wess and B.~Zumino, Phys. Lett. B 37 (1971) 95.
\hfil\break
  \ E.~Witten, Nucl. Phys. B 223 (1983) 422.
\hfil\break
  \ W.A.~Bardeen, Phys. Rev. 184 (1969) 1848.
\hfil\break
  \ifx\answ\bigans
  \ K.~Fujikawa, Phys. Rev. Lett. 42 (1979) 1195; Phys. Rev.
  D 21 (1980) 2848;
\hfil\break\indent Phys. Rev. D 22 (1980) 1499; Phys. Rev. D 23 (1981) 2262.
   \else
   \ K.~Fujikawa, Phys. Rev. Lett. 42 (1979) 1195;
    \hfil\break\indent Phys. Rev. D 21 (1980) 2848;
    \hfil\break\indent Phys. Rev. D 22 (1980) 1499;
    \hfil\break\indent Phys. Rev. D 23 (1981) 2262.\fi
\hfil\break
  \ifx\answ\bigans
  \ B.~Zumino, in: Relativity, Groups, and Topology II, eds.
  B.S.~DeWitt and R.~Stora 
    \hfil\break\indent (North-Holland, Amsterdam, 1984) p.1291.
   \else
   \ B.~Zumino, in: Relativity, Groups, and Topology II, eds. B.S.~DeWitt
    \hfil\break\indent and R.~Stora 
	(North-Holland, Amsterdam, 1984) p.1291.\fi
\hfil\break
  \ B.~Zumino, Y.-S.~Wu and A.~Zee, Nucl. Phys. B 239 (1984) 477.
\hfil\break
  \ W.A.~Bardeen and B.~Zumino, Nucl. Phys. B 244 (1984) 421.
\hfil\break
  \ L.~Alvarez-Gaum\'e and P.~Ginsparg, Ann. Phys. 161 (1985) 423.
\hfil\break
  \ N.K.~Pak and P.~Rossi, Nucl. Phys. B 250 (1985) 279.}\ 
which for the case at hand is
\eqn\EQxix{\vcenter{\openup2\jot\halign{$\displaystyle#\hfil$&
     $\displaystyle\hfil#$\cr
    {\cal L}_{WZ} =
     {1\over16\pi^{2}} \varepsilon^{\mu\nu\rho\sigma}
     &\left\{{\bf B}_{\mu}\dn{\bf B}_{\rho} \Tr\left\lgroup
      {\sigma^{3}\over2}\left(\boldSigma^{\dagger}\left({\bf W}_{\sigma}
       + \Ls\right)\boldSigma - {\bf W}_{\sigma}\right)
       \right\rgroup\right.\cr
     &\left.
       + \dm{\bf B}_{\nu} \Tr\left\lgroup{\bf W}_{\rho}\Ls\right\rgroup
       - {1\over3}{\bf B}_{\mu}
            \Tr\left\lgroup\Ln\Lr\Ls\right\rgroup\right\},\cr
}}}
with $\Lm = \left(\dm\boldSigma\right)\boldSigma^{\dagger}$, ${\bf W}_{\mu} =  -
ig_{2}W^{i}_{\mu}{\sigma^{i}\over2}$ and $ {\bf B}_{\mu} = - ig_{1}B_{\mu}$.

Thus, our complete low-energy Lagrangian is:
\eqn\EQxx{\vcenter{\openup2\jot\halign{$\displaystyle\hfil#$
                                &$\displaystyle#\hfil$\cr
 \ifx\answ\bigans
 {\cal L} = &- {1\over4}W^{i\mu\nu}W^{i}_{\mu\nu}
    - {1\over4}B^{\mu\nu}B_{\mu\nu}
      + {F_{\pi}^{2}\over4}\Tr\left\lgroup {\cal A}^{\mu}
                                    {\cal A}_{\mu}\right\rgroup
      + {\cal L}_{gf} + \overline l_{L} i\Dsl l_{L}
      + \overline e_{R} i\Dsl e_{R}\cr
      &\hskip 5em + \overline Q\left(i\calDsl^{Q} - m_{Q}\right)Q
   + \overline Q\calAsl\gamma_{5}Q + {\cal L}_{WZ}+\dots,\cr
 \else
 {\cal L} = &- {1\over4}W^{i\mu\nu}W^{i}_{\mu\nu}
    - {1\over4}B^{\mu\nu}B_{\mu\nu}
      + {F_{\pi}^{2}\over4}\Tr\left\lgroup {\cal A}^{\mu}
                                    {\cal A}_{\mu}\right\rgroup
      + {\cal L}_{gf}\cr
    & + \overline l_{L} i\Dsl l_{L}
      + \overline e_{R} i\Dsl e_{R}
      + \overline Q\left(i\calDsl^{Q} - m_{Q}\right)Q
   + \overline Q\calAsl\gamma_{5}Q + {\cal L}_{WZ}+\dots,\cr\fi
}}}
where ${\cal L}_{gf}$ contains the gauge fixing and Fadeev-Popov ghost
terms necessary for gauge field quantization.  (We assume, of course,
that electroweak gauge fixing is accomplished through a
``$\xi$-gauge'' condition,\ref\FLS{
  K.~Fujikawa, B.W.~Lee and A.I.~Sanda, Phys. Rev. D 6 (1972) 2923.}\ 
${\cal L}_{gf} =
{1\over2\xi}\left(\dm A^{\mu} - \xi M\pi\right)^{2}$, in which we know
how to prove the equivalence theorem.) The quarks now have derivative
couplings to the pions and thus the VVP triangles connecting pions to
vector currents in the $q$ basis are replaced by Vector -- Vector --
Axial vector (VVA) triangles and gauge -- Goldstone contact
interactions in the Wess-Zumino term.  The VVA triangles, and all
other terms explicitly involving quarks, decouple as the quark mass is
taken to infinity.  The Wess-Zumino term does not explicitly involve
the quarks and does not decouple.  It cannot decouple because it is
not gauge invariant.  In the $q$ basis, the quarks cannot decouple
because they must cancel the anomalous gauge dependence of the
leptons.  In the $Q$ basis, the Wess-Zumino term balances the gauge
dependence of the leptons; the $Q$ quarks are gauge invariant and
there is no obstacle to their decoupling.  We can directly integrate
out the quarks in the $Q$ basis but not in the $q$ basis.

\newsec{The Equivalence Theorem}

The simplest process involving the anomaly which one might investigate
is the well known decay $\pi^{0} \longrightarrow \gamma\gamma$, and
its analog through the equivalence theorem $Z_{L} \longrightarrow
\gamma\gamma$.  This process, however, is not an appropriate test of
the equivalence theorem which is explicitly formulated to be valid for
high energy processes ($E \gg M_{W,Z}$), and relies upon the fact that
the longitudinal polarization vector for fast moving gauge bosons is
approximately proportional to the momentum of the particle.  This
approximation is clearly invalid in the rest frame.

Let us therefore examine a process where we can produce energetic
gauge bosons such as $\overline\nu + \nu \longrightarrow Z_{L} +
\gamma$.  This process provides a particularly clean example to study
as the more realistic process $e^{+} + e^{-} \longrightarrow Z_{L} +
\gamma$ is also more complicated because of  $Z$ -- $\gamma$
interference.

The relevant diagrams for the $q$ and $Q$ bases are shown in figures 
\fig\figone{Feynman diagrams for $Z_{L}\gamma$ and $\pi\gamma$ production 
with quarks in the $q$ basis.  The Dirac structure of the external
$Z_{L}$\ /$\pi$ couplings to the fermion triangles is indicated.  The full
vertex factors are: \figone a) $\mp{ig_{2}\over2\cos\theta_{W}}
\epsl(p)\gamma_{5} \approx \mp{i\over F_{\pi}}\psl\gamma_{5}$ for
$\left(u\atop d\right)$ quarks; \figone b) $\approx{i\over F_{\pi}}
\psl\gamma_{5}$; and \figone c) $\pm{m_{Q}\over F_{\pi}}\gamma_{5}$.}\ and 
\fig\figtwo{Feynman diagrams for $Z_{L}\gamma$ and $\pi\gamma$ production 
with quarks in the $Q$ basis.  The Dirac structure of the external
$Z_{L}$\ /$\pi$ couplings to the fermion triangles is indicated.  The full
vertex factors are: \figtwo a) $\mp{ig_{2}\over2\cos\theta_{W}}
\epsl(p)\gamma_{5} \approx \pm{i\over F_{\pi}}\psl\gamma_{5}$ for
$\left(u\atop d\right)$ quarks; \figtwo b) $\approx{i\over F_{\pi}}\psl
\gamma_{5}$; and \figtwo c) $\pm{1\over F_{\pi}}\psl\gamma_{5}$.  Figure
\figtwo d) represents $\pi\gamma$ production via the Wess-Zumino
term.}\ respectively.
(Implicitly included are the triangle diagrams with the
fermions propagating in the opposite direction around the loop.)
In the $q$ basis, there is no Wess-Zumino term, so
the VVP triangle of figure \figone c completely describes $\pi^{0}\gamma$
production, while in the $Q$ basis $\pi^{0}\gamma$ production results
from the sum of the VVA triangle of Figure \figtwo c and the Wess-Zumino
interaction of Figure \figtwo d. The amplitude for $\pi^{0} \gamma$
production via the Wess-Zumino interaction is
\eqn\EQxxi{{\cal M}^{WZ}_{\pi\gamma} = {ieg_{2}^{2}\over64\pi^{2}F_{\pi}} {\left(1 -
    4\sin^{2}\theta_{W}\right)\over\cos^{2}\theta_{W}}
    {\overline\nu\gamma_{\lambda}\left(1-\gamma_{5}\right)\nu\over
     s-M_{Z}^{2}}\varepsilon^{\alpha\beta\nu\lambda} p_{1\alpha} p_{2\beta}
     \epsilon^{\gamma}_{\nu}(p_{2}),}
where $\epsilon^{\gamma}$ is the photon polarization vector.  The
other amplitudes, computed in the limit that
$\epsilon_{\mu}^{Z_{L}}(p_{1}) \approx {p_{1\mu}\over M_{Z}} + {\cal
O}\left({M_{Z}\over E}\right)$, are:
\eqn\EQxxii{\vcenter{\openup2\jot\halign{$\displaystyle\hfil#$
                               &$\displaystyle#\hfil$\cr
  {\cal M}^{Q}_{\pi\gamma} &= -{\cal M}^{WZ}_{\pi\gamma}\left[1
       - 2\int_{0}^{1}\mkern-10mu dx\mkern-5mu\int_{0}^{1-x}\mkern-30mu dy
        {m_{Q}^{2}\over{\left(y^{2}-y\right)M_{Z}^{2}
         - xy\left(s-M_{Z}^{2}\right) + m_{Q}^{2}}}\right]\cr
  {\cal M}^{q}_{\pi\gamma} &= {\cal M}^{WZ}_{\pi\gamma}
       + {\cal M}^{Q}_{\pi\gamma}\cr
  {\cal M}^{l}_{Z\gmma} &= -i{\cal M}^{WZ}_{\pi\gamma}
       + {\cal O}\left({M_{Z}\over E}\right)\cr
   {\cal M}^{Q}_{Z\gmma}&= -i{\cal M}^{Q}_{\pi\gamma}
       + {\cal O}\left({M_{Z}\over E}\right) = {\cal M}^{q}_{Z\gmma}\cr
}}}
where $x$ and $y$ are Feynman parameters.  Up to a phase,
\eqn\EQxxiii{{\cal M}^{Q}_{\pi\gamma} + {\cal M}^{WZ}_{\pi\gamma} = {\cal
  M}^{q}_{\pi\gamma} = {\cal M}^{l}_{Z\gmma} + {\cal M}^{Q}_{Z\gmma}
  + {\cal O}\left({M_{Z}\over E}\right),}
exactly as the equivalence theorem asserts.

There are several features worth noticing.  First, by explicit calculation,
${\cal M}^{Q}_{\pi\gamma} + {\cal   M}^{WZ}_{\pi\gamma}$ is exactly equal
to ${\cal M}^{q}_{\pi\gamma}$.  This is as it must be since the $q$ and $Q$
formulations are equivalent.  Second, the validity of the
equivalence theorem is independent of the constituent quark mass
$m_{Q}$.  We see this most clearly in the $Q$ basis where we find that
gauge boson production via quark triangles is equivalent to would-be
Goldstone production via quark triangles, and gauge boson production via
lepton triangles is equivalent to would-be Goldstone production via the
Wess-Zumino interaction.  The quark triangles for gauge and would-be
Goldstone boson production have identical mass dependence while the lepton
triangles and the Wess-Zumino interaction are independent of the quark mass.

Finally, for $M^{2}_{Z} \ll s \ll m^{2}_{Q}$, the quark triangles, ${\cal
M}^{Q}_{Z\gmma}$ and ${\cal M}^{Q}_{\pi\gamma}$ are suppressed relative to
the lepton triangles ${\cal M}^{l}_{Z\gmma}$ and the Wess-Zumino term ${\cal
M}^{WZ}_{\pi\gamma}$ respectively, providing nonvanishing production
amplitudes for $Z_{L}\gamma$ and $\pi^{0}\gamma$.  For $s \gg m^{2}_{Q}$,
however, the quark triangles are not suppressed relative to the lepton
triangles and the Wess-Zumino term and tend to cancel them so that the
production amplitudes are quite small.  For example, at an intermediate
energy $M^{2}_{Z} \ll s \ll m^{2}_{Q}$, the cross section for $Z_{L}\gamma$
production is approximately $s$ independent, 
\eqn\EQxxiv{\sigma_{Z_{L}\gmma}\left(M^{2}_{Z} \ll s \ll m^{2}_{Q}\right) =
   {\alpha\alpha^{2}_{W}\left(1-4\sin^{2}\theta_{W}\right)\over
      3072\pi^{2}F_{\pi}^{2}\cos^{4}\theta_{W}}
       \left(1-{s\over6m^{2}_{Q}}\right),}
while at high energy, $s \gg m^{2}_{Q}$, the cross section falls
rapidly with s,
\eqn\EQxxv{\sigma_{Z_{L}\gmma}\left(s \gg m^{2}_{Q}\right) =
   {\alpha\alpha^{2}_{W}\left(1-4\sin^{2}\theta_{W}\right)\over
      3072\pi^{2}F_{\pi}^{2}\cos^{4}\theta_{W}}{\pi^{4}m_{Q}^{4}\over s^{2}}
       \left(1 + {1\over\pi^{2}}\left(\ln{s\over m_{Q}^{2}}\right)^{2}
         \right)^{2}.}

Let us now return to reference~\DT.  Donoghue and Tandean compute
$\pi^{0}\gamma$ production exclusively through the Wess-Zumino interaction,
while finding that $Z_{L}\gamma$ production vanishes because the quark
triangles exactly cancel the lepton triangles.  From
equation~\EQxxii, we see that their computation of $\pi^{0}\gamma$
production is a valid low energy (and/or heavy quark) approximation, ($s \ll
m^{2}_{Q}$ so that $\left|{\cal M}^{Q}_{\pi\gamma} \right| \ll \left|{\cal M
}^{WZ}_{\pi\gamma}\right|$) while their result for $Z_{L}\gamma$ production
is a valid high energy (and/or light quark) approximation. ($s \gg
m^{2}_{Q}$ so that ${\cal M}^{Q}_{Z\gmma} \approx -{\cal M}^{l}_{Z\gmma}$)
Each is valid within a particular energy regime, but they are never
simultaneously valid.

\newsec{Comments}

One complication that we did not include, though we know it to be
present in the real world, is that the strong QCD forces renormalize
the hadronic part of the axial vector current so that the $\overline
Q\calAsl\gamma_{5}Q$ term in equation~\EQxx\ should enter with a
coefficient $g_{A}$ which could be determined experimentally.  The
introduction of $g_{A}$ is the only change that needs to be made to
the Lagrangian in the $Q$ basis.  Both ${\cal M}^{Q}_{Z\gmma}$ and ${\cal
M}^{Q}_{\pi\gamma}$ are multiplied by $g_{A}$ while the lepton couplings and
the Wess-Zumino interaction are unchanged so that the equivalence theorem
remains intact. More complicated modifications must be made in the $q$
basis,\MM\ but the conclusions are of course the same.

Another complication that we have not addressed involves enlarging the
scalar sector.  This would happen if more than one electroweak doublet
were to participate in the condensate.  In this case, we would again find
the equivalence theorem to be valid.  As discussed above, in the $Q$ basis
quark triangle production of high energy gauge bosons is manifestly
equivalent to quark triangle production of would-be Goldstone bosons.
Since the strength of the Wess-Zumino term is determined by the strength
of the quark anomaly\WZ\ which, because of anomaly cancellation is equal
in magnitude to the lepton anomaly, there is again equivalence between
gauge boson production via lepton triangles and would-be Goldstone production
via the Wess-Zumino interaction. Thus, the overall equivalence is maintained.

\newsec{Conclusions}

The equivalence theorem is a direct consequence of the BRS symmetry of
a spontaneously broken gauge theory.  Since BRS symmetry is the special
form of the classical gauge symmetry which survives the quantization
procedure, the equivalence theorem is a necessary consequence of gauge
invariance.  In this letter, we have constructed an explicit example of
how the equivalence theorem is satisfied even in the presence of an
anomalous symmetry breaking condensate.  Gauge invariance demands that
the anomaly associated with the condensate be cancelled by some other
part of the theory.  Together, the two parts combine to preserve gauge
invariance and in so doing guarantee the equivalence theorem.

In demonstrating the validity of the equivalence theorem, we have seen
that anomalous condensate models permit gauge boson production through
fermion triangle diagrams.  It may be that there are `anomalous
technicolor' models for which this production mechanism has observable
consequences.\ref\WKb{W.~Kilgore, in preparation.}

\bigskip\bigskip\noindent{\bf Acknowledgements}
\bigskip\noindent
I would like to thank Marcus Luty, Greg Keaton and Michael Chanowitz for
many helpful discussions.

\bigskip\noindent\DOE

\listrefs\listfigs
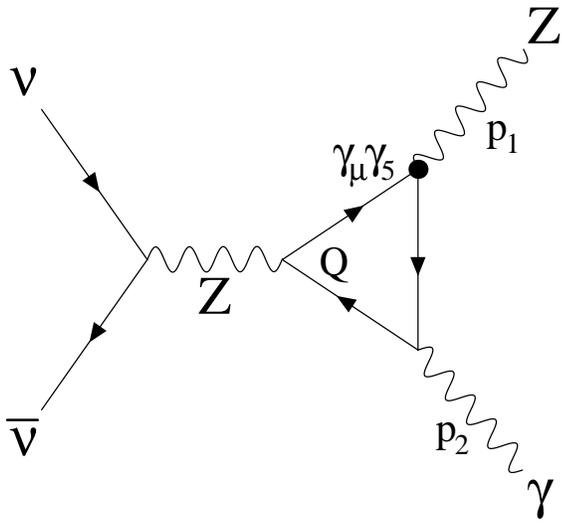
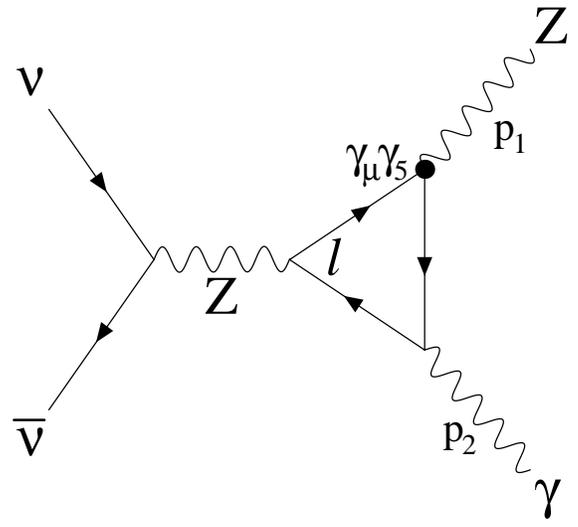
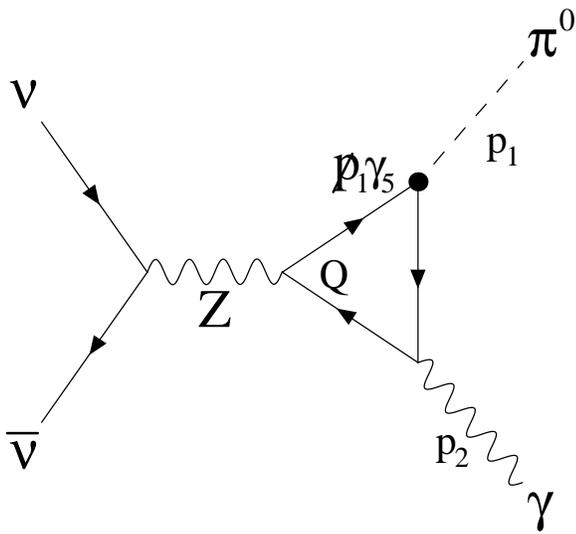
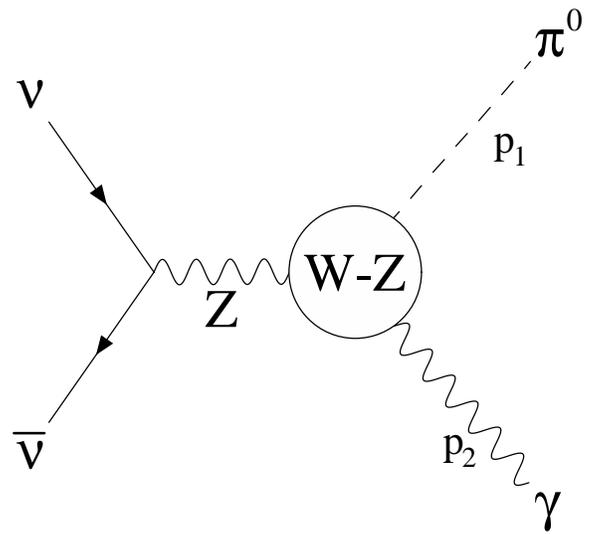
\end